\documentstyle[12pt]{article}
\begin {document}

\begin{flushright}
OITS-599\\
April 1996\\
\end{flushright}
 
\vskip.5cm
\begin{center}
{\Large \bf Fluctuations and Entropy Indices \\of QCD Parton Showers}
\vskip .75cm
  {\bf  Zhen CAO and Rudolph C. HWA}
\vskip.5cm
 {Institute of
Theoretical Science and Department of Physics\\ 
University of Oregon,
Eugene, OR 97403}
\end{center}

\vskip.2cm

\begin{abstract}

The branching processes in parton showers are studied in perturbative QCD for both quark and gluon
jets. The emphasis is on the nature of fluctuations of both the parton multiplicities and the spatial
patterns of the final states.  Effective measures of such fluctuations are calculated from the data
obtained by Monte Carlo simulations.  The entropy indices are used to characterize chaoticity.  Both
running and fixed couplings are considered.  The fixed coupling case is used to study the onset of
chaos.  Implications of the results are discussed.
\end{abstract}

\section{\bf Introduction}

The possibility of chaotic behavior of particle production in branching processes has recently been
investigated \cite{1,2} with emphasis on the search for appropriate measures of chaoticity. 
Unlike classical nonlinear dynamics where the Lyapunov exponent $\lambda$ can be defined to
characterize divergent distances between trajectories, no such conventional description of chaotic
behavior of branching processes exist.  One of the measures found in \cite{1,2} is the
exponent
$\kappa$ characterizing the exponential dependence of the normalized multiplicity variance on the
average multiplicity; another is the entropy index $\mu_q$.  With those two measures it is possible to
distinguish the properties of particle production of non-Abelian from Abelian dynamics to the extent
that one may regard the former as chaotic and the latter not.

There are two aspects of that investigation that need further extension and exploration.  The first
is an obvious one:  the pure gauge theory of gluons only in the perturbative QCD branching should be
extended to include quarks.  The other is less obvious.  If the QCD dynamics is indeed chaotic,
there is a question rooted in the conventional chaos theory that should be asked.  That is, at what
value of the control parameter can one identify the onset of chaos?  In QCD there is no externally
controllable parameter with which one can tune the dynamics.  Since the nonlinearity of the
dynamics that gives rise to branching can be represented perturbatively as a self-coupling term
whose strength is parametrized by the strong coupling constant $\alpha_s$, we may regard
$\alpha_s$ as the control parameter.  Thus in order to investigate the nature of the onset of
chaos, we shall reconsider the QCD branching problem by treating $\alpha_s$ as fixed, and then
vary it by hand to study the dependences of $\kappa$ and $\mu_q$ on $\alpha_s$.

In Sec.\ 2 we treat the full QCD problem with quarks and gluons and with running $\alpha_s
(q^2)$.  In Sec.\ 3 the onset of chaos is studied by varying the fixed $\alpha_s$.  

\section{QCD Parton Showers}

As in \cite{1,2}, we follow Odorico's procedure \cite{3} to develop the algorithm for Monte
Carlo simulation of parton shower in QCD.  We shall not repeat here the details of the algorithm
already described in \cite{2}.  What is new now is the inclusion of quarks in the branching
processes, and the consideration of quark jets in addition to the gluon jets.

The splitting functions are
\begin{eqnarray}
P_{G\rightarrow GG} (z)=2 N_c \left[\frac{1-z}{z} + \frac{z}{1-z} + z(1-z) \right] \quad,
\label{1}
\end{eqnarray}    
\begin{eqnarray}
P_{G\rightarrow q \overline{q}} (z) = \frac{1}{2}\left[z^2 +(1-z)^2 \right] \quad,
\label{2}
\end{eqnarray}
\begin{eqnarray}
P_{q \rightarrow qG}(z) = \frac{N^2_c -1}{2N_c} \cdot \frac{1+z^2}{1-z} \quad,
\label{3}
\end{eqnarray}
and the running coupling constant is 
\begin{eqnarray}
\alpha_s (q^2) = \frac{12 \pi}{11\, N_c - 2N_f} \cdot \frac{1}{\ell n(q^2/\Lambda^2)} \quad .
\label{4}
\end{eqnarray}
We shall choose massless quark number $N_f =3$, color number $N_c =3$, and set the QCD scale at
$\Lambda = 250$ MeV.  The approximations made to treat the infrared and collinear divergences result
in the Sudakov form factors \cite{3,4} that have the following forms for the gluon and quark
(or antiquark) vertices, respectively,
\begin{eqnarray}
\Delta_G (Q^2, K^2) = {\rm exp} \left\{
 -\frac{2}{9}\,\ell n\left(
 \frac{\ell n Q^2/\Lambda^2}{\ell n
K^2/\Lambda^2}\right)
 \int_{\epsilon}^{1-\epsilon}dz \left[
 P_{G \rightarrow GG}(z) +3P_{G\rightarrow
q \overline{q}} (z) \right]
 \right\}\quad,
\label{5}
\end{eqnarray}
\begin{eqnarray}
\Delta_q (Q^2, K^2) = {\rm exp} \left\{-\frac{2}{9}\,\ell n \left(\frac{\ell n Q^2/\Lambda^2}{\ell n
K^2/\Lambda^2}\right) \int_{\epsilon}^{1-\epsilon}dz  P_{q \rightarrow qG}(z)   \right\}\quad,
\label{6}
\end{eqnarray}
where $\epsilon= Q_0^2 /Q^2$.  We take the end of branching to occur when $q^2 \leq Q_0^2 =1
{\rm GeV}^2$.

Since a gluon can now go into a number of channels, the simulation of a branching into resolvable
partons requires the designation of a specific final state.  To that end we calculate the ratio
\begin{eqnarray}
R=\frac{\int_{\epsilon}^{1-\epsilon} P_{G \rightarrow GG}(z) dz}
{\int_{\epsilon}^{1-\epsilon}
\left[P_{G \rightarrow GG}(z) +3 P_{G \rightarrow q \overline{q}}(z) \right]dz}\quad,
\label{7}
\end{eqnarray}
and compare it with a random number $R_{\#} \in [0,1]$.  In $R_{\#} <R$, then the $G\rightarrow GG$
channel is chosen.  Because of the divergence of (\ref{1}) at $z=0$ and $1$, $R$ is very close to
$1$; it is $\approx 0.8$ even for $\epsilon =0.3$.  Thus the self-reproduction of gluons is a far more
dominant process than $q \overline{q}$ pair creation.  By this fact alone we do not expect the gluon
jet to change too much by the inclusion of quarks.  However, the quark jets are very different from
the gluon jets, as we shall show.

We first show in Fig.\ 1 our result on the distribution of maximum number of generations,
$P(i_{max})$, for both the gluon-initiated shower (G jet) and quark-initiated shower (Q jet)
at two different values of initial virtuality $Q$.  The distributions are obtained after $10^5$
samples are simulated.  There are less generations of branching in a Q jet than in a G jet because
$P_{q
\rightarrow qG} (z)$ has divergence only at $z=1$ (soft gluon limit), while $P_{G \rightarrow GG}
(z)$ diverges at both $z=0$ and $1$.  Thus the highly reproductive gluons are generally produced
with low momenta (and consequently low virtualities), resulting in less multiplicities
for the Q jet.  For the purpose of studying multiplicity fluctuations at different generations of
branching, it is meaningful to consider only the subset of all showers that have the same maximum
generations, $i_{max}$.  We shall choose $i_{max}$ to be at the peaks of $P(i_{max})$.  More
specifically, we set $i_{max}= 6$ and $11$ for Q and G jets, respectively, at $Q/Q_0 =10^2$, and
$i_{max}=11$ and $22$ at $Q/Q_0 =10^3$. 

For each fixed $i \leq i_{max}$ there is a distribution of multiplicities $n_i$ at that $i$.  That
distribution $P(n_i)$ is shown in Fig.\ 2 for the case of $Q/Q_0 =10^3$ and for the values $i=6$ and
$11$.  It is clear that $\left< n_i \right>$ for G jet is greater than that for Q jet for both $i$
values.  The fluctuations from $\left< n_i \right>$ are measured by the normalized variance
\begin{eqnarray}
V_i = \frac{\left< n_i^2 \right>- \left< n_i \right>^2}{\left< n_i \right>^2} \quad .
\label{8}
\end{eqnarray}
In Fig.\ 3 are shown $V_i$ vs $\left< n_i \right>$ for the various cases of jet type and $Q/Q_0$. 
The most striking revelation one sees in that figure is that the general features of the Q and G
jets do not differ by very much.  Focusing on $Q/Q_0 =10^3$ we see that their $V_i$ values reach
about the same peak value even though the corresponding $\left< n_i \right>$ are different for the
Q and G jets.  That small difference, however, accounts for a difference in the value of $\kappa$
defined by 
\begin{eqnarray}
V_i \propto \left< n_i \right>^{\kappa}\quad.
\label{9}
\end{eqnarray} 
Straightline fits to the linear portions for the log-log plots for the $Q/Q_0 =10^3$ case yield
\begin{eqnarray}
\kappa =  \begin{array}{ll} 0.62 ,&(Q \,\, \mbox{jet},\quad 10< \left< n_i \right> <50)\quad,\\
        0.29 ,&(G \,\mbox{jet},\quad 20< \left< n_i \right> <70)\quad.\end{array}
\label{10}
\end{eqnarray}
This difference in the values of $\kappa$ distinguishes the Q jet from the G jet.  It is a
quantitative measure of the differences exemplified by the thin and thick solid lines in Fig.\ 2. 
The dotted lines in Fig.\ 3 represent the results of pure gauge theory with $N_f =0$.  Note that
they are indistinguishable from the new results for G jet with quarks included.

The above result may be regarded as the characteristics of the temporal behaviors of the branching
processes, when $\left< n_i \right>$ is interpreted as an analogue of the time elapsed in a
classical tragectory \cite{2}.  For a more effective description of the fluctuations, we now examine
the spatial patterns in phase space at the end of branching and study how they fluctuate from event to
event.  In terms of the final-state momentum fraction $x$ of a parton, where $x=\Pi_i z_i$, the
inclusive distribution $\rho(x)$ is highly nonuniform in the interval $0<x<1$.  Even in terms of the
$\zeta$ variable, where $\zeta =-{\rm log}_{10} x$, the distributions appear Gaussian, as shown in
Fig.\ 4.  Apart from the heights of $dn/d \zeta$ it appears that there is little difference
between the Q and G jets.  However, those are only single-particle distributions.  To see spatial
patterns it is necessary to study the normalized factorial moments $F_q$ in small bins.  To that
end we must first use a spatial variable $X$ in terms of which $\rho(X)$ is uniform.  Adopting the
cumulative variable \cite{1,2,5,6} defined by 
\begin{eqnarray}
X(\zeta) =\int_{\zeta_1}^{\zeta} \rho(\zeta^\prime) d \zeta^\prime / \int_{\zeta_1}^{\zeta_2}
\rho(\zeta^\prime) d
\zeta^\prime \quad , 
\label{11}
\end{eqnarray}    
where $\zeta_{1,2}$ are the extrema of the $\zeta$ range in Fig.\ 4, we have $\rho (X)=$ constant
for $0 \leq X \leq 1$.  It is this unit interval in $X$ space that we partition into $M$ bins of
width $\delta =1/M$.

For an event $e$ let $n_j$ be the multiplicity of partons in bin $j$ at the end of branching so that
the factorial moment for that event is 
\begin{eqnarray}
f_q^e (M) =M^{-1} \sum_{j=1}^{M} n_j (n_j-1) \cdot \cdot \cdot (n_j -q+1) \quad . 
\label{12}
\end{eqnarray}
The spatial pattern for the $e$th event at resolution $\delta$ is then described by 
\begin{eqnarray}
F_q^e = f_q^e/(f_1^e)^q  \quad. 
\label{13}
\end{eqnarray}  
The fluctuation of $F_q^e$ from event to event is a dynamical property of QCD branching that results
in the loss of information about the final state.  Thus it is important to quantify the degree of
that fluctuation, which we do by taking the moments of $F_q^e$:
\begin{eqnarray}
C_{p,q} = \left<(F_q^e)^p\right>/\left<F_q^e\right>^p \quad , 
\label{14}
\end{eqnarray}
where $\left<\cdots \right>$ means averaging over all events.  The exponent $p$ can be any real
number, not necessarily an integer. Since $F_q^e$ deviates more from $1$ when $\delta$ is smaller, we
search for the power-law behavior in $M$, i.\ e.\ , 
\begin{eqnarray}
C_{p,q}(M) \propto M^{\psi_q (p)} \quad . 
\label{15}
\end{eqnarray}
The entropy index is defined by \cite{1,2}
\begin{eqnarray}
\mu_q =\left. \frac{d}{dp}\psi_q(p) \right| _{p=1}  \quad   . 
\label{16}
\end{eqnarray}

The simulated data we use for this part of the analysis are not restricted, as was done for the
study of $V_i$ where the maximum number of generations, $i_{max}$, was chosen at the peak of
$P(i_{max})$.  In the study of spatial patterns that change from event to event, we consider all
events whatever $i_{max}$ may be.  After simulating $10^5$ events we calculate $C_{p,q} (M)$, the
results for which are shown in Fig.\ 5 for Q and G jets at $Q/Q_0 =10^3$.  Evidently, those moments
are much larger for Q jet than for G jet; thus the former has much larger spatial fluctuations.  The
entropy indices provide a more efficient way of describing that property, as shown by the plots in
Fig.\ 6.  The values of $\mu_q$ for the G jet are not significantly different from the ones
determined in \cite{2}, where $N_f =0$.  But they are much smaller than $\mu_q$ for the Q jet.  This
difference is one of the important findings in this work.  It could not have been inferred by
examining the single-particle distributions such as those shown in Fig.\ 4.  Because of the
differences between $P_{q \rightarrow qG}$ and $P_{G \rightarrow GG}$, the evolution, the
multiplicity, and the distribution of $F_q$ are all different for the Q jet, as compared to the G
jet.  If the branching dynamics for the G jet is chaotic, then it is even more so for the Q jet.  

\section{Varying the Fixed Coupling}

As we have found in \cite{1,2}, the non-Abelian QCD dynamics is different from the Abelian
$\chi$ model in that $\kappa >0$ and $\mu_q$ are larger.  Whether or not it means that the former is
indeed chaotic is not certain, since we lack a definitive criterion for chaoticity in problems where
no trajectory can be defined unambiguously and where the number of degrees of freedom increases with
evolution.  One way of investigating further the nature of possible chaotic behavior in QCD is to
examine the onset of chaos.  Since $\alpha_s$ controls the strength of the nonlinearity in the
problem, it is reasonable to propose that $\alpha_s$ be varied in order to find the threshold of the
chaotic behavior.  Of course, it means that we must first fix $\alpha_s$ and calculate the
fluctuation properties of the parton showers, and then examine how those properties depend on
$\alpha_s$.

With $\alpha_s$ not running, the Sudakov form factors in (\ref{5}) and (\ref{6}) lose one degree of
log each and become 
\begin{eqnarray}
\Delta_G (Q^2, K^2)= {\rm exp} \left\{-\frac{\alpha_s}{2 \pi} \ell n \frac{Q^2}{K^2}
\int_{\epsilon}^{1-
\epsilon}dz \left[P_{G \rightarrow GG}(z)+ 3 P_{G \rightarrow q \overline{q}}(z) \right] \right\}
\quad ,
\label{17}
\end{eqnarray}
\begin{eqnarray}
\Delta_q (Q^2, K^2)= {\rm exp} \left\{-\frac{\alpha_s}{2 \pi} \ell n \frac{Q^2}{K^2}
\int_{\epsilon}^{1-
\epsilon}dz\, P_{q \rightarrow qG}(z) \right\} \quad .
\label{18}
\end{eqnarray}
Apart from this change the simulation of parton showers is just as in Sec.\ 2.  We shall fix $Q/Q_0$
at $10^3$, and let $\alpha_s$ vary over the range $0.05$ to $0.3$, which roughly covers the running
range as the virtuality degrades from $Q$ to $Q_0$.  At the low end of $\alpha_s$ (near $0.05$) the
corresponding $Q^2$ value is unrealistic high, but is nevertheless considered here because of our
interest in the problem at the very small $\alpha_s$.

For fixed $\alpha_s$ the distributions of $i_{max}$ have shapes that are similar to those in
Fig.\ 1 for running $\alpha_s$.  The peak of $P(i_{max})$, denoted by $i_{max}^{peak}$, increases
with $\alpha_s$ for both Q and G jets, as shown in Fig.\ 7.  It is important to recognize that the
number of generations in a branching process is very small at low $\alpha_s$.  That is because the
survivability of a parton without emitting a resolvable gluon or creating a $q \overline{q}$ pair is
nearly $1$ at very small $\alpha_s$, as is evident in (\ref {17}) and (\ref {18}).  That is quite
unlike the running $\alpha_s(q^2)$ case, where $\alpha_s(q^2)$ always gets bigger as $q^2$ evolves
toward $Q_0^2$.  The smallness of $i_{max}^{peak}$ at small $\alpha_s$ is particularly acute for the
Q jet, since when a gluon is emitted by a quark, the gluon momentum and virtuality are suppressed,
thus depriving the Q jet a rich source of parton generations.  This property of low multiplicity at
low $\alpha_s$ will be recalled later to explain other features to be uncovered.

From the simulated events we calculate $V_i$ as before and find the dependence on $\left<n_i \right>$
as shown in Fig.\ 8.  Evidently, there is a universality in how $V_i$ depends on $\left<n_i
\right>$.  The increase of
$\alpha_s$ merely extends the range.  There is a bend in the rise of $V_i$ for G jet, but the rise
seems to persist for the Q jet, a phenomenon that we have already encountered in Fig.\ 3.  From
Fig.\ 8 we gain no insight on the development of chaos, if indeed the temporal behavior is chaotic,
since there is no value of $\alpha_s$ that we can identify as the onset of chaos.

Next, for the fluctuations of spatial patterns we show $C_{p,q}(M)$ in Figs.\ 9 and 10 for Q and G
jets, respectively.  The general features are similar to Fig.\ 5 for running $\alpha_s(q^2)$, but
there are important differences.  It helps to focus on a fixed $p$ and examine the dependence on
$\alpha_s$.  While the scaling behaviors for $\alpha_s = 0.15$ and $0.25$ are rather similar, in
some cases even hard to distinguish, the behavior for $\alpha_s = 0.05$ stands out markedly
different.  That is, $|\psi_q(p)|$ is much larger at small $\alpha_s$ than at larger $\alpha_s$. 
This feature is most pronounced for G jet at $q=2$, where the only curves significantly different
from $0$ are for $\alpha_s = 0.05$.  Comparing the vertical scales of the two figures, one can also
see that the Q jet has larger magnitudes of log $C_{p,q}$ than the G jet.  These features are both
related to the fact that the parton multiplicity is low for small $\alpha_s$ and especially so for
the Q jet.  Fluctuations are usually large when the averages are small.

The characteristics of Figs.\ 9 and 10 are represented in capsule form by the entropy indices
$\mu_q$.  In Fig.\ 11 we show $\mu_q$ for various values of $\alpha_s$, while in Fig.\ 12 the
dependences on $\alpha_s$ are shown for fixed $q$.  For $\alpha_s>0.15$, $\mu_q$ approaches
independence on $\alpha_s$, but as $\alpha_s$ decreases toward $0$, $\mu_q$ increases sharply.  This
behavior has not been anticipated, and is opposite to what is expected if $\alpha_s$ is to play the
role of a control parameter that can turn on chaotic behavior as it is increased from a small enough
value.

As we have noted earlier, the parton multiplicities are low, when $\alpha_s$ is small, so their
fluctuations relative to their averages are very large.  That is what $\mu_q$ measures.  Since chaos
implies unpredictability of the final state of a parton shower, our results indicate that a parton
shower is more chaotic at low $\alpha_s$ than at high $\alpha_s$.  Thus our conclusion here is that
$\alpha_s$ cannot be treated as a control parameter, useful for tuning out chaos.  Perturbative QCD
is such an intricate nonlinear dynamics that it cannot be rendered approximately linear by
decreasing the strength of the nonlinear term.  Without the nonlinear coupling there is no
evolution, no particle production, and no linear dynamics.  

\section {Conclusions}

Our study of the full QCD branching problem that has both quarks and gluons has revealed several
properties of the parton showers.  The gluon jet with quarks included is not very different from
the case of pure gauge theory \cite {1,2} . But the quark jet exhibits more
fluctuations (relative to the average) than the gluon jet.  The measures used to quantify the
fluctuations are $V_i$ for the temporal behavior and $\mu_q$ for the spatial behavior.  They both
possess features that are characertistic of chaotic behavior.

In the fixed coupling case we have found that the results on fluctuations are roughly the same as in
running $\alpha_s$ problem, provided that the fixed $\alpha_s$ is not set at a very low value,
outside the range of the running $\alpha_s$.  However, in attempting to learn about the onset of
chaos we have treated $\alpha_s$ as a control parameter and examined the parton showers with
$\alpha_s$ being allowed to approach a very small value.  The hope was to see how the chaotic
behavior would grow as $\alpha_s$ is increased. What we have found is that the QCD branching
processes are more chaotic at very low $\alpha_s$.  Unlike classical nonlinear dynamics, the
nonlinearity in the non-Abelian gauge theory cannot be turned off without causing the system to
lose the mechanism for temporal evolution through branching.  A small amount of nonlinearity
leads to large fluctuations relative to the mean number of partons produced, resulting in more
unpredictable final states.

The measure by which these properties can be described is the entropy index $\mu_q$.  We have
seen how $\mu_q$ has emerged as a highly effective description of the degree of fluctuations of
spatial patterns.  It is our opinion that the method of analysis employed here is not only
suitable for the study of parton showers, but also for all problems that involve changing
spatial patterns, including the conventional problems in classical chaos.    

\begin{center}
\subsection*{Acknowledgment}
\end{center}

This work was supported in
part by the U.S. Department of Energy under Grant No. DE-FG06-91ER40637.

\newpage
\centerline {\bf Figure Captions}

\begin{enumerate}
\item[Fig.  1{\quad}]Distributions of the maximum number of generations in QCD branching processes for
quark ($Q$) and gluon ($G$) jets.

\item[Fig.  2{\quad}]Multiplicity distributions at $i=6$ and $11$ for quark and gluon jets.

\item[Fig.  3{\quad}] Normalized variance vs average multiplicity as $i$ is increased from $1$ to
$i_{max}$.  The dotted lines are for pure gauge theory without quarks.

\item[Fig.  4{\quad}]Single-particle distribution in $\zeta = -{\rm log} x$.

\item[Fig.  5{\quad}]Log-log plots of $C_{p,q}$ vs $M$ for various values of $p$ and $q$.

\item[Fig.  6{\quad}]Entropy indices $\mu_q$ for quark and gluon jets.

\item[Fig.  7{\quad}]The dependence of $i_{max}^{peak}$ on the fixed coupling $\alpha_s$, where
$i_{max}^{peak}$ is the value of $i_{max}$ at the peak of the distribution $P(i_{max})$.  

\item[Fig.  8{\quad}]$V_i$ vs $\left< n_i \right>$ for various values of fixed $\alpha_s$. 

\item[Fig.  9{\quad}]Log-log plots of $C_{p,q}$ vs $M$ for quark jets only when $\alpha_s$ is set at
three possible values.

\item[Fig.  10{\quad}]Same as Fig.\ 9 for gluon jets.

\item[Fig.  11{\quad}]$\mu_q$ vs $q$ for various $\alpha_s$.

\item[Fig.  12{\quad}]$\mu_q$ vs $\alpha_s$ for various $q$.

\end{enumerate}

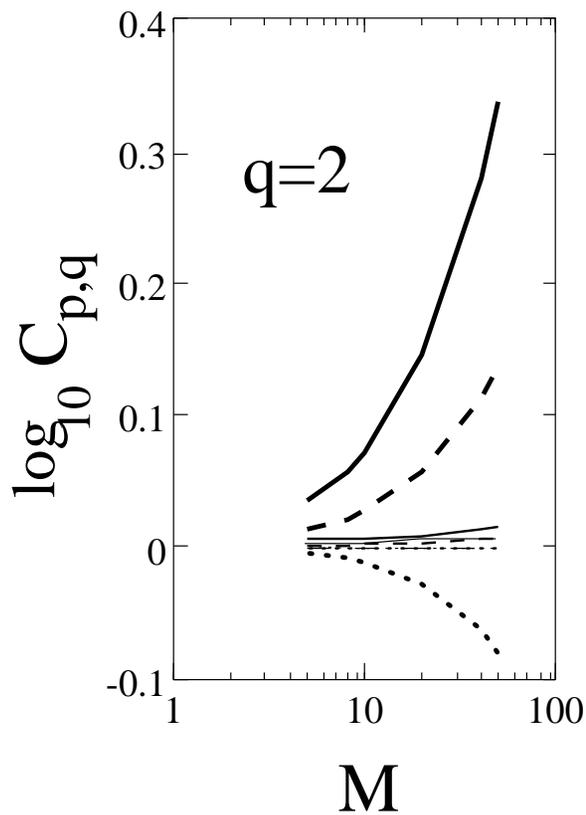
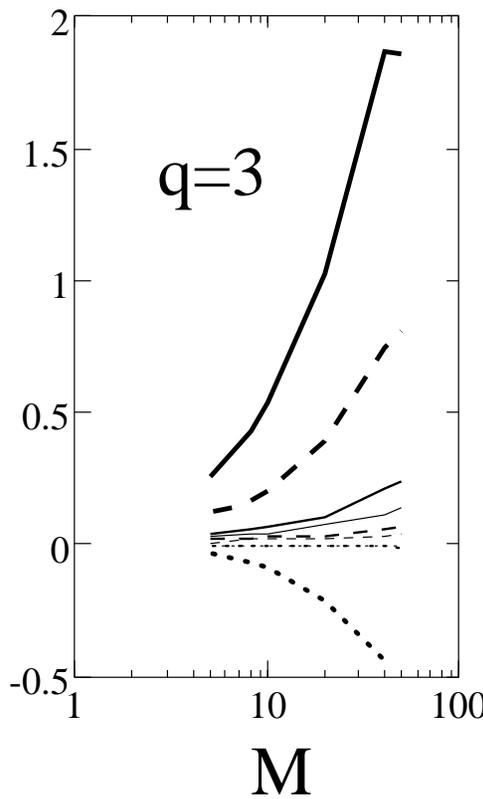
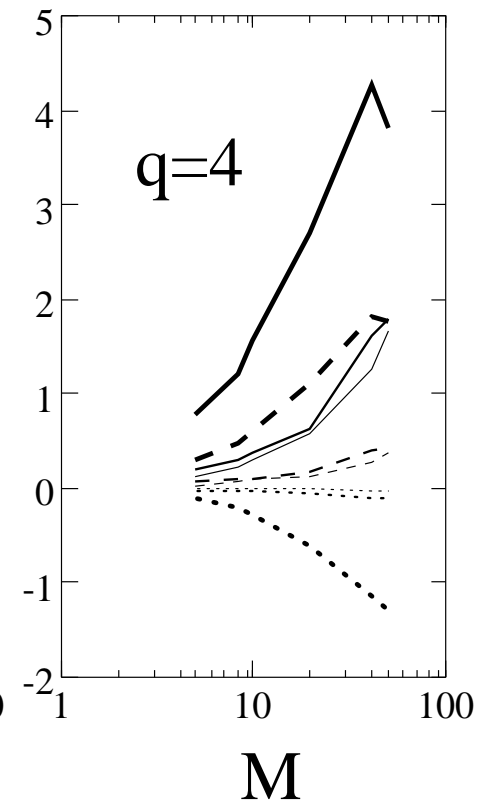
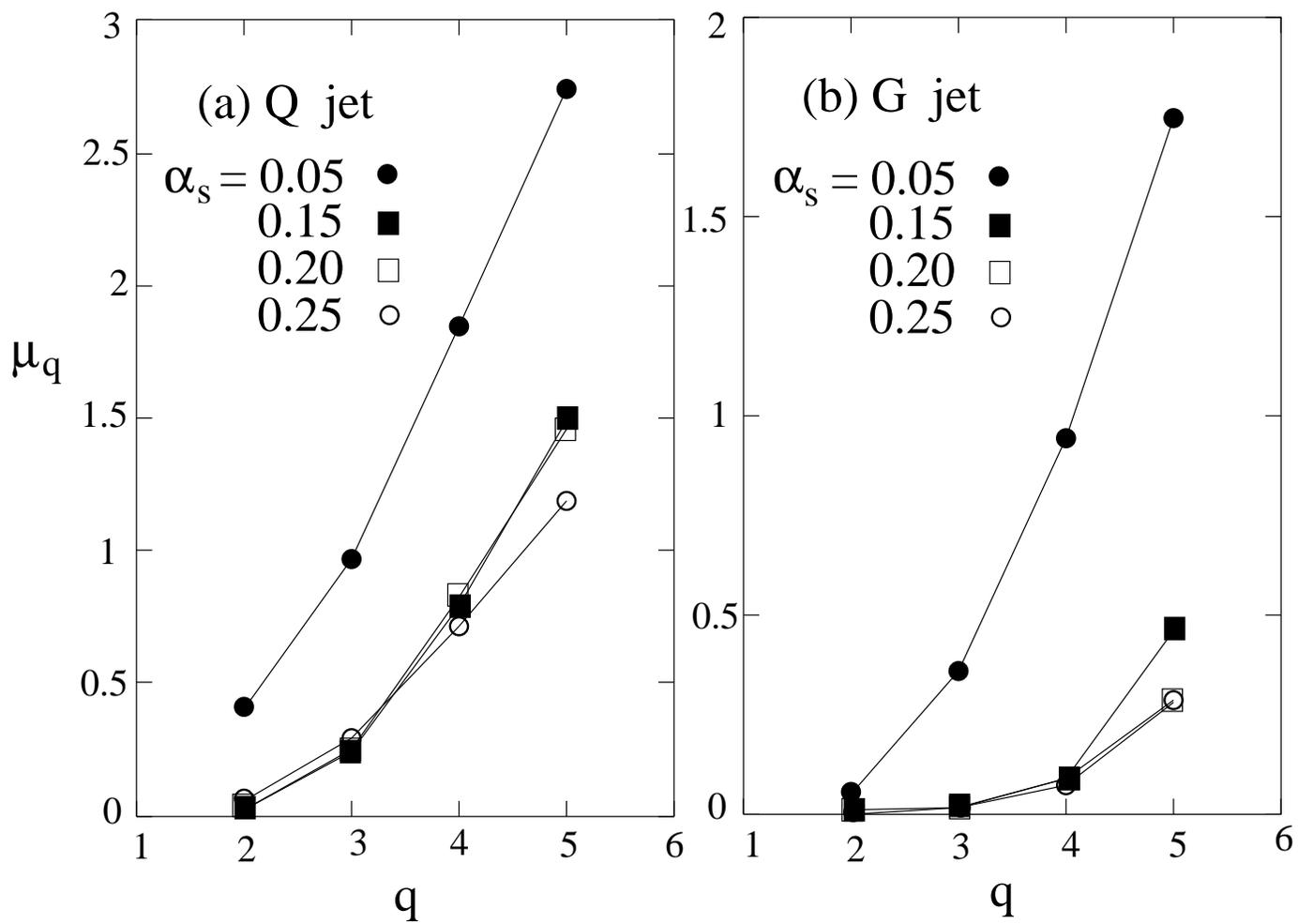
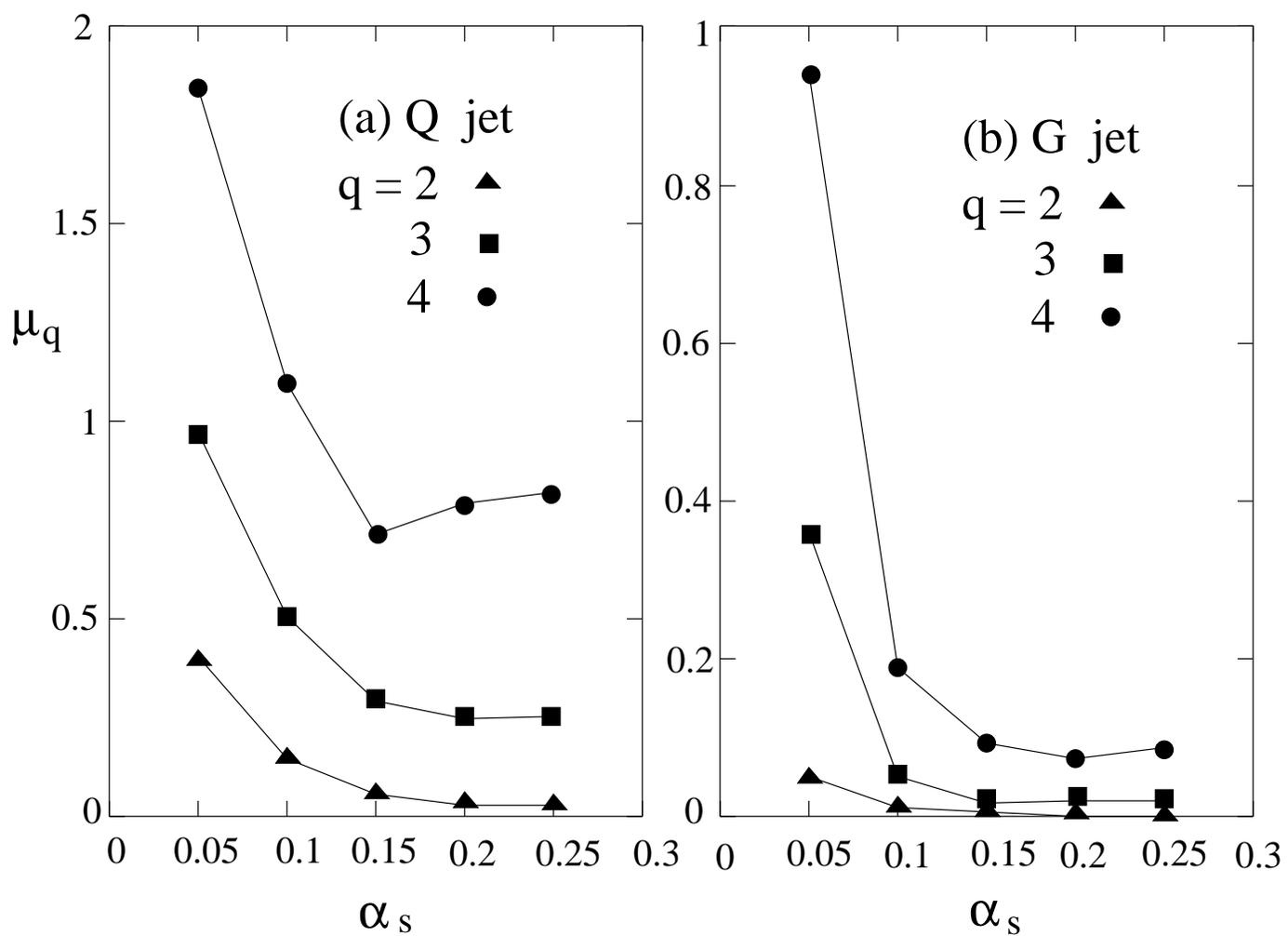
\end{document}